\documentclass[pra,10pt,reprint,superscriptaddress,showpacs]{revtex4-1}

\usepackage{epsfig}
\usepackage{amsmath}
\usepackage{graphicx}
\usepackage{graphics}
\usepackage{float}
\usepackage{amssymb}
\usepackage[usenames,dvipsnames]{color}
\usepackage{natbib}
\usepackage[outdir=./]{epstopdf}
\usepackage{verbatim}
\usepackage{wrapfig}
\usepackage{subfigure}
\usepackage{bbold}
\usepackage{dsfont}

\usepackage{mathtools}

\newcommand{\bra}[1]{\left\langle #1\right|}
\newcommand{\ket}[1]{\left| #1\right\rangle}

\newcommand{\opav}[3]{\langle #1 | #2 | #3 \rangle}

\newcommand{\beq}{\begin{equation}}
\newcommand{\eeq}{\end{equation}}

\newcommand{\tr}{\text{Tr}}
\newcommand{\tot}[1]{#1_{\text{tot}}}
\newcommand{\re}{\text{Re}}

\begin{document}

\title{A general framework for the Quantum Zeno and anti-Zeno effects}

\author{Adam Zaman Chaudhry}
\email{adam.zaman@lums.edu.pk}
\affiliation{School of Science \& Engineering, Lahore University of Management Sciences (LUMS), Opposite Sector U, D.H.A, Lahore 54792, Pakistan}

\begin{abstract}

A general treatment of the quantum Zeno and anti-Zeno effects is presented which is valid for an arbitrary system-environment model in the weak system-environment coupling regime. It is shown that the effective lifetime of a quantum state that is subjected to repeated projective measurements depends on the overlap of the spectral density of the environment and a generalized `filter function' which in turn depends on the system state that is repeatedly being prepared, the measurement interval, the system parameters, the system-environment coupling, and the environment correlation function. This general framework is then used to study explicitly the Zeno to anti-Zeno crossover behavior for the spin-boson model where a single two-level system is coupled to a bosonic environment. It is shown that our framework reproduces results for the usual population decay case as well as for the pure dephasing model, while at the same time allowing us to study the Zeno to anti-Zeno transition when both decay and dephasing take place. These results are then extended to many two-level systems coupled collectively to the bosonic environment to show that the distinction between the Zeno and anti-Zeno effects becomes more pronounced as the number of two-level systems increases.

\end{abstract}

\date{\today}

\maketitle

\section{Introduction}

It is well-known that if a quantum system is subjected to rapidly repeated measurements, the temporal evolution of the quantum system slows down \cite{Sudarshan1977}. This effect, known as the Quantum Zeno effect (QZE), has attracted widespread interest both theoretically and experimentally due to its relevance to the foundations of quantum mechanics as well as possible applications in quantum technologies \cite{FacchiPhysLettA2000,FacchiPRL2002, FacchiJPA2008, WangPRA2008, ManiscalcoPRL2008, FacchiJPA2010, MilitelloPRA2011, RaimondPRA2012, SmerziPRL2012, WangPRL2013, McCuskerPRL2013, StannigelPRL2014, ZhuPRL2014, SchafferNatCommun2014,SignolesNaturePhysics2014,DebierrePRA2015,AlexanderPRA2015,QiuSciRep2015,Slichterarxiv2015}. Surprisingly, however, it has also been found that if the measurements are not rapid enough, the opposite effect can occur, that is, the measurements can actually accelerate quantum transitions. This effect has been appropriately dubbed the quantum anti-Zeno effect (QAZE) \cite{KurizkiNature2000, KoshinoPhysRep2005}. Since this discovery, both the QZE and QAZE have been studied in many different physical setups such as nanomechanical oscillators \cite{BennettPRB2010}, Josephson junctions \cite{BaronePRL2004}, disordered spin systems \cite{YamamotoPRA2010}, and localized atomic systems \cite{RaizenPRL2001}. Generally speaking, the focus has been to study the population decay of quantum systems, whereby the quantum system is prepared in an excited state, and the system is thereafter repeatedly checked to see if the system is still in the excited state or not \cite{KurizkiNature2000, KoshinoPhysRep2005,BennettPRB2010,BaronePRL2004,YamamotoPRA2010,RaizenPRL2001,ManiscalcoPRL2006,SegalPRA2007,ZhengPRL2008,
AiPRA2010,ThilagamJMP2010,ThilagamJCP2013}. It has then been shown that the decay rate depends on the overlap of the spectral density of the environment and a measurement-induced level width \cite{KurizkiNature2000}. This overlap changes as the measurement rate changes, generally leading to a crossover from the Zeno regime with large measurement rates to the anti-Zeno regime with relatively smaller measurement rates. 

With increasingly sophisticated quantum technologies, it becomes important to investigate what happens beyond such population decay models. For instance, we can envisage repeatedly preparing not simply the excited state, but rather a superposition state of the ground and excited states. Such a superposition state will undergo not only population decay but also dephasing \cite{Schlosshauerbook}. To this end, the QZE and the QAZE have been studied for a pure dephasing model in Ref.~\cite{ChaudhryPRA2014zeno}. It was found therein that there are considerable differences between this case and the population decay case. For example, for the pure dephasing model in the weak coupling regime, the lifetime of the quantum state depends on the overlap of the spectral density of the environment and a `filter function' that is different from the usual sinc-squared function obtained for the population decay model. Moreover, by examining the collective dephasing of many two-level systems, it was found that multiple Zeno and anti-Zeno regimes can be found. At the same time, however, these results are limited in the sense that they are only applicable for exactly solvable pure dephasing models. Nevertheless, these differences motivate us to investigate the QZE and the QAZE for more arbitrary system-environment Hamiltonians and state preparations. To this end, in this work, we derive an expression for the effective lifetime of a quantum state, subjected to repeated projective measurements, that is valid for weak system-environment coupling strength. This effective lifetime depends on the overlap between the spectral density of the environment and a generalized `filter function' that in turn depends on the state that is repeatedly prepared, the environment correlation function, the system Hamiltonian parameters, the measurement interval, and the system-environment coupling. This expression is general in the sense that no assumption is made beforehand about the actual form of the system-environment Hamiltonian or the state that is being repeatedly prepared. 

Once we have derived a general expression for the effective lifetime of a quantum state, our main task then is to actually evaluate the generalized filter function. To illustrate the application of our formalism, the task of evaluating this filter function is first performed for the usual population decay model to show that we reproduce the well-known sinc-squared filter function. Similarly, we produce results that are in agreement with the exactly solvable pure dephasing model in the weak-coupling regime. We then consider the QZE and QAZE for the general spin-boson model where generally both population decay and dephasing take place. In particular, we show that the filter function is different for the general case as compared with the pure dephasing and decay cases and must be carefully evaluated. We then calculate the effective lifetime of the quantum state, and thereby the Zeno to anti-Zeno crossover behavior, to further illustrate the differences as compared to the pure dephasing and decay cases. We next consider a collection of two-level systems collectively coupled to an environment that causes both dephasing and decay. In this case, we find that the effective decay rate is amplified depending on the number of particles coupled to the environment, thereby making the use of the correct filter function even more important. 

This paper is organized as follows. In Sec.~II, we derive our general expression for the effective lifetime of the quantum state. In Sec.~III, this formalism is applied to the spin-boson model with pure dephasing, population decay, as well as the case with both decay and dephasing present at the same time. This study of the QZE and the QAZE is also performed for the so-called large spin-boson model which models the behavior of many two-level systems collectively coupled to an environment. We conclude in Sec.~IV. Some details of the calculations can be found in the Appendix. 

\section{The general framework}

We start by considering a general system-environment Hamiltonian written in the form  
\begin{equation*}
H = H_S + H_B + V, 
\end{equation*}
where $H_S$ is the system Hamiltonian, $H_B$ is the Hamiltonian of the environment, and $V$ describes the system-environment interaction. At time $t = 0$, we prepare the initial system state $\ket{\psi}$. $N$ repeated projective measurements, given by the projector $P_\psi = \ket{\psi}\bra{\psi}$, are now applied with equal time interval $\tau$. It should be noted that before each measurement, we apply the operator $e^{iH_S\tau}$ \cite{MatsuzakiPRB2010,ChaudhryPRA2014zeno} in order to remove the system evolution due to $H_S$ itself \footnote{In other words, we consider the so-called `dynamical fidelity' to characterize the effective decay rate.}. If the survival probability of the quantum state is $S$ after $N$ time intervals, it is convenient to write $S \equiv e^{-\Gamma(\tau)N\tau}$, thereby allowing us to interpret $1/\Gamma(\tau)$ as the effective lifetime of the quantum state. Alternatively, $\Gamma(\tau)$ can be interpreted as the effective decay rate of the quantum state. Since $S = [s(\tau)]^N$ \cite{noteoncorrelations}, where $s(\tau)$ is the survival probability associated with one measurement, we find that the effective decay rate is $\Gamma(\tau) = -\frac{1}{\tau} \ln s(\tau)$. Thus, our primary objective is to find $s(\tau)$ for the system-environment Hamiltonian, provided that we start from the pure system state $\ket{\psi}$. This is what we now proceed to do.
	
	Let us first find an expression for $\rho(\tau)$, the state of the system at time $t = \tau$, just before the system is measured. Now, the total system-environment state at this time is 
		$$ \rho_{\text{tot}}(\tau) = U(\tau) \rho_{\text{tot}}(0)U^\dagger(\tau), $$
		where $U(\tau)$ represents the unitary time-evolution due to the total system-environment Hamiltonian $H$. We can then find the state of the system at time $\tau$ as 
		$$ \rho_S(\tau) = \text{Tr}_B \lbrace U(\tau) \rho_{\text{tot}}(0)U^\dagger(\tau) \rbrace, $$
		where $\text{Tr}_B$ denotes taking partial trace over the environment. Generally speaking, however, it is usually impossible to find $U(\tau)$ exactly. Instead, we resort to perturbation theory, which can be used if we assume that the system-environment is weak. We write $U(\tau)$ as $U(\tau) = U_0(\tau)U_I(\tau)$, where $U_0(\tau) = e^{-i(H_S + H_B)\tau}$ is the `free' unitary time-evolution operator, and $U_I(\tau)$ can be expanded as a perturbation series, that is, $U_I(\tau) = \mathds{1} + A_1 +  A_2 + \hdots$, where $A_1$ and $A_2$ are the first and second order corrections respectively. Note also that we set $\hbar = 1$ throughout this work. The system density matrix at time $\tau$ is then 
		\begin{align}
		\rho_S(\tau) \approx \, &\text{Tr}_B \lbrace U_0(\tau) [\rho_{\text{tot}}(0) + \rho_{\text{tot}}(0) A_1^\dagger + \rho_{\text{tot}}(0)A_2^\dagger \notag \\
								&+ A_1 \rho_{\text{tot}}(0) + A_2 \rho_{\text{tot}}(0) + A_1 \rho_{\text{tot}}(0) A_1^\dagger] U_0^\dagger(\tau) \rbrace,
		\label{simplifythis}
		\end{align}
		correct to second order in the system-environment coupling strength.
		
		To proceed further, we assume that the initial system-environment state can be represented as $\rho_{\text{tot}} = \rho_S(0) \otimes \rho_B$, that is, the initial system-environment state is a simple product state \footnote{This is a rather standard assumption in the open quantum systems literature. However, the validity of this assumption is questionable when the system-environment coupling strength is weak and/or the number of particles collectively coupled with the environment is large [see, for instance, A.~Z.~Chaudhry and J.~B.~Gong, Phys.~Rev.~A, \textbf{87}, 012129 (2013); \textbf{88}, 052107 (2013); Can.~J.~Chem.~\textbf{92}, 119 (2014).]}. Moreover, we also assume that we can write $V = F \otimes B$, where $F (B)$ is an operator belonging to the system (environment) Hilbert space. The more general case where $V = \sum_\mu F_\mu \otimes B_\mu$ can be dealt with via a straightforward extension. The operators $A_1$ and $A_2$ introduced above are then found to be 
		\begin{align*}
		A_1 &= -i\int_0^\tau \widetilde{V}(t_1) dt_1, \\
		A_2 &= -\int_0^\tau dt_1 \int_0^{t_1} dt_2 \widetilde{V}(t_1)\widetilde{V}(t_2),
		\end{align*}
		with $\widetilde{V}(t) \equiv U_0^\dagger(t) V U_0(t) = U_S^\dagger(t)FU_S(t) \otimes U_B^\dagger(t) B U_B(t) = \widetilde{F}(t)\widetilde{B}(t)$ , where $U_S(t) = e^{-iH_St}$ and $U_B(t) = e^{-iH_Bt}$. We now simplify Eq.~\eqref{simplifythis} term by term. First, we find that 
			$$ \text{Tr}_B \lbrace U_0(\tau) \rho_{\text{tot}}(0)U_0^\dagger(\tau)\rbrace = \widetilde{\rho}_S(\tau), $$
			where $\widetilde{\rho}_S(\tau) = U_S(\tau)\rho_S(0)U_S^\dagger(\tau)$ is the system density matrix if the system and environment are not coupled together. Next, we find that 
			\begin{align*}
			&\text{Tr}_B \lbrace U_0(\tau) \rho_{\text{tot}}(0)A_1^\dagger U_0^\dagger(\tau)\rbrace = \\
			&i \int_0^\tau dt_1 \, U_S(\tau) \rho_S(0) \widetilde{F}(t_1) U_S^\dagger(\tau) \text{Tr}_B \lbrace U_B(\tau) \rho_B \widetilde{B}(t_1) U_B^\dagger(\tau)\rbrace.
			\end{align*}
			But $\text{Tr}_B \lbrace U_B(\tau) \rho_B \widetilde{B}(t_1) U_B^\dagger(\tau)\rbrace = \tr_B \lbrace \rho_B \widetilde{B}(t_1)\rbrace$, which is zero for the system-environment models usually considered \cite{BPbook}. Similarly, 
			$$ \text{Tr}_B \lbrace U_0(\tau) A_1 \rho_{\text{tot}}(0) U_0^\dagger(\tau)\rbrace = 0. $$
			Next, we find that 
			\begin{multline*} 
			\tr_B \lbrace U_0(\tau) A_2 \tot{\rho}(0) U_0^\dagger (\tau) \rbrace = \\
 -\int_0^\tau dt_1 \int_0^{t_1} dt_2 \, U_S(\tau) \widetilde{F}(t_1) \widetilde{F}(t_2) \rho_S(0) U_S^\dagger(\tau) C(t_1,t_2),
			\end{multline*}
			with the environment correlation function $C(t_1,t_2)$ defined as $C(t_1,t_2) = \langle \widetilde{B}(t_1)\widetilde{B}(t_2)\rangle_B = \tr_B \lbrace \widetilde{B}(t_1)\widetilde{B}(t_2)\rho_B\rbrace$. Similarly, 
			\begin{multline*}
			\tr_B \lbrace U_0(\tau) \tot{\rho}(0) A_2^\dagger U_0^\dagger (\tau) \rbrace = \\
 -\int_0^\tau dt_1 \int_0^{t_1} dt_2 \, U_S(\tau) \rho_S(0) \widetilde{F}(t_2) \widetilde{F}(t_1) U_S^\dagger(\tau) C(t_2,t_1).
			\end{multline*}	
			Finally, 
			\begin{multline*}
			\tr_B\lbrace U_0(\tau)A_1 \tot{\rho}A_1^\dagger U_0^\dagger(\tau)\rbrace = \\
			\int_0^\tau dt_1 \int_0^\tau dt_2 U_S(\tau) \widetilde{F}(t_1)\rho_S(0)\widetilde{F}(t_2)U_S^\dagger(\tau) C(t_2,t_1).
			\end{multline*}
			Using the fact that $\int_0^\tau dt_1 \int_0^\tau dt_2 = \int_0^\tau dt_1 \int_0^{t_1} dt_2 + \int_0^\tau dt_2 \int_0^{t_2} dt_1$, we can manipulate further to obtain
			\begin{align*}
			&\tr_B\lbrace U_0(\tau)A_1 \tot{\rho}A_1^\dagger U_0^\dagger(\tau)\rbrace = \\
			&\int_0^\tau dt_1 \int_0^{t_1} dt_2 U_S(\tau) \widetilde{F}(t_1)\rho_S(0)\widetilde{F}(t_2)U_S^\dagger(\tau) \\
			&\times \langle \widetilde{B}(t_2)\widetilde{B}(t_1)\rangle_B + \text{h.c.},
			\end{align*}
			where h.c. denotes hermitian conjugate. Putting all the terms back together, the system density matrix can be written as 
			\begin{align*}
			&\rho_S(\tau) = U_S(\tau) \biggl( \rho_S(0) + \int_0^\tau dt_1 \int_0^{t_1} dt_2 \bigl\lbrace \langle \widetilde{B}(t_1)\widetilde{B}(t_2) \rangle_B \notag \\
			&\times [\widetilde{F}(t_2)\rho_S(0),\widetilde{F}(t_1)] + \text{h.c.}\bigr\rbrace \biggr) U_S^\dagger (\tau).
			\end{align*}
			We can simplify this further by noting that the environment correlation function $C(t_1,t_2) = \langle \widetilde{B}(t_1)\widetilde{B}(t_2) \rangle_B$ generally depends on the time difference $t_1 - t_2$ only. This then motivates us to introduce $t' = t_1 - t_2$. The system density matrix at time $\tau$ then becomes 
			\begin{align}
			&\rho_S(\tau) = U_S(\tau) \biggl( \rho_S(0) + \int_0^\tau dt_1 \int_0^{t_1} dt' \bigl\lbrace C(t') \notag \\
			&\times [\widetilde{F}(t_1 - t')\rho_S(0),\widetilde{F}(t_1)] + \text{h.c.}\bigr\rbrace \biggr) U_S^\dagger (\tau).
			\label{rdmtimetau}
			\end{align}
			with the simplified notation $C(t') = \langle \widetilde{B}(t')\widetilde{B}(0) \rangle_B$ for the environment correlation function.
			
			Once we have the expression for the density matrix at time $\tau$, we can compute the survival probability. This survival probability can be simply calculated as one minus the probability of getting some result other than the state $\ket{\psi}$ that we have started off with. Consequently, it is useful to define the projection operator $P_{\psi^\perp}$ that projects onto the subspace orthogonal to the state $\ket{\psi}$. Moreover, we must also keep in mind that, just before performing the measurement, we perform a unitary operator (which is implemented on a very short time-scale) in order to remove the evolution due to the system Hamiltonian itself. This unitary operator then removes the $U_S(\tau)$ and $U_S^\dagger(\tau)$ that can be found to the left and right of the right hand side of Eq.~\eqref{rdmtimetau}. Thus, we can write the survival probability as 
			\begin{align*}
			s(\tau) = &1 - \int_0^\tau dt_1 \int_0^{t_1} dt' \biggl( C(t') \notag \\
			&\times \tr \bigl\lbrace P_{\psi^\perp}[\widetilde{F}(t_1 - t')\rho_S(0),\widetilde{F}(t_1)] + \text{h.c.}\bigr\rbrace \biggr),
			\end{align*}
			where $\tr$ denotes simply taking the trace over the system only, and we have used the fact that $\tr\lbrace\rho_S(0)P_{\psi^\perp}\rbrace = 0$. Using the fact that for any operator $X$, $\tr(X + X^\dagger) = 2 \re[\tr(X)]$, where $\re$ denotes taking the real part, we further simplify to 
			\begin{align*}
			s(\tau) = &1 - 2 \,\re \biggl[ \int_0^\tau dt \int_0^{t} dt' \biggl( C(t') \notag \\
			&\times \tr \bigl\lbrace P_{\psi^\perp}[\widetilde{F}(t - t')\rho_S(0),\widetilde{F}(t)]\bigr\rbrace \biggr)\biggr],
			\end{align*}
			and we have replaced $t_1$ with $t$ for notational simplicity. Since $\tr\lbrace P_{\psi^\perp} \widetilde{F}(t) \widetilde{F}(t - t')\rho_S(0)\rbrace = 0$, we get
			\begin{align*}
			s(\tau) = &1 - 2 \,\re \biggl[ \int_0^\tau dt \int_0^{t} dt' \biggl( C(t') \notag \\
			&\times \tr \bigl\lbrace P_{\psi^\perp} \widetilde{F}(t - t')\rho_S(0) \widetilde{F}(t) \bigr\rbrace \biggr)\biggr].
			\end{align*}
			Now, the environment correlation function will typically be of the form $C(t') = \sum_k |g_k|^2 f(\omega_k,t')$, where $g_k$ is the coupling strength of the system with mode $k$ of the environment, and $f(\omega_k,t')$ is simply a function containing the remaining information about $C(t')$. This sum is usually replaced by an integral over the frequencies of the environment via the substitution $\sum_k |g_k|^2 (\hdots) \rightarrow \int_0^\infty d\omega \,J(\omega) (\hdots)$, thereby introducing the spectral density $J(\omega)$ of the environment. For now, let us simply note that 
			\begin{equation*}
			s(\tau) = 1 - \sum_k |g_k|^2 Q'(\omega_k,\tau), 
			\end{equation*}
			where 
			\begin{align*}
			Q'(\omega_k,\tau) = &2 \,\re \biggl[ \int_0^\tau dt \int_0^{t} dt' \biggl( f(\omega_k,t') \notag \\
			&\times \tr \bigl\lbrace P_{\psi^\perp} \widetilde{F}(t - t')\rho_S(0) \widetilde{F}(t) \bigr\rbrace \biggr)\biggr].
			\end{align*}			
			What we are really interested in is the effective decay rate $\Gamma(\tau) = -\frac{1}{\tau} \ln s(\tau)$. For weak coupling strength, the deviation of $s(\tau)$ from unity is small. Therefore, the effective decay rate can be approximated as 
			\begin{equation}
			\Gamma(\tau) = \sum_k |g_k|^2 Q(\omega_k,\tau), 
			\end{equation}
			where 
			\begin{align}
			Q(\omega_k,\tau) = &\frac{2}{\tau} \,\re \biggl[ \int_0^\tau dt \int_0^{t} dt' \biggl( f(\omega_k,t') \notag \\
			&\times \tr \bigl\lbrace P_{\psi^\perp} \widetilde{F}(t - t')\rho_S(0) \widetilde{F}(t) \bigr\rbrace \biggr)\biggr].
			\end{align}
			By considering the environment to be made up of a continuum of frequencies, and thus introducing the spectral density, we can write instead
			\begin{equation}
			\label{basicformula}
			\Gamma(\tau) = \int_0^\infty  d\omega \, J(\omega) Q(\omega,\tau), 
			\end{equation}
			where the generalized `filter function' $Q(\omega,\tau)$ is
			\begin{align}
			Q(\omega,\tau) = &\frac{2}{\tau} \,\re \biggl[ \int_0^\tau dt \int_0^{t} dt' \biggl( f(\omega,t') \notag \\
			&\times \tr \bigl\lbrace P_{\psi^\perp} \widetilde{F}(t - t')\rho_S(0) \widetilde{F}(t) \bigr\rbrace \biggr) \biggr].
			\end{align}	
			In other words, if the system-environment coupling is weak, the effective decay rate of a quantum state $\Gamma(\tau)$ depends on the overlap of the spectral density of the environment $J(\omega)$ and an effective filter function	$Q(\omega,\tau)$. This filter function depends on the frequency of the measurement, the state that is repeatedly prepared, the way that the system is coupled to the environment, and part of the environment correlation function. In the next section, we will evaluate this effective filter function for a few common system-environment models. At this point, let us note that the behavior of $\Gamma(\tau)$ as a function of $\tau$ allows us to identify the Zeno and anti-Zeno regimes. Namely, when $\Gamma(\tau)$ is an increasing function of $\tau$, we are in the Zeno regime, while if $\Gamma(\tau)$ is a decreasing function of $\tau$, then we are in the anti-Zeno regime \footnote{It is worth emphasizing that we are using a `local' approach to QZE-QAZE transitions \cite{KurizkiNature2000, SegalPRA2007, ThilagamJMP2010}. An alternative approach is to compare the measurement modified decay rate with the `free' decay rate, that is, the decay rate without measurement [see, for instance, P.~Facchi, H.~Nakazato and S.~Pascazio, Phys.~Rev.~Lett.~{\textbf{86}}, 2699 (2001)].  In this work, we use the former approach.}. We also note that for the more generalized system-environment coupling $V = \sum_\mu F_\mu \otimes B_\mu$,  we find that $\Gamma(\tau)$ is again given by Eq.~\eqref{basicformula}, but now the filter function is 
			\begin{align}
			Q(\omega,\tau) = &\frac{2}{\tau} \,\re \biggl[ \sum_{\mu\nu} \int_0^\tau dt \int_0^{t} dt' \biggl( f_{\mu\nu} (\omega,t') \notag \\
			&\times \tr \bigl\lbrace P_{\psi^\perp} \widetilde{F}_\nu(t - t')\rho_S(0) \widetilde{F}_\mu(t) \bigr\rbrace \biggr) \biggr],
			\end{align}	
			 where $f(\omega,t')$ is extracted from $C_{\mu\nu}(t') = \int_0^\infty d\omega \, J(\omega) f_{\mu\nu}(\omega,t') = \langle \widetilde{B}_\mu(t') \widetilde{B}_\nu(0)\rangle_B$. 
			 
	\section{Application to the spin-boson model}
	
	\subsection{The usual population decay model}
	
	To illustrate the formalism that we have developed, let us start from the system-environment Hamiltonian 
	$$ H = \frac{\varepsilon}{2}\sigma_z + \sum_k \omega_k b_k^\dagger b_k +  \sum_k (\sigma^+ g_k^* b_k + \sigma^- g_k b_k^\dagger). $$
	Here a two-level system with level spacing $\varepsilon$ interacts with an environment, which is modeled as a collection of harmonic oscillators, and undergoes population decay. $\sigma_z$ is the standard Pauli matrix, $\sigma^+$ and $\sigma^-$ are the raising and lowering operators respectively, $\omega_k$ are the frequencies of the environment oscillators, $b_k^\dagger$ and $b_k$ are the creation and annihilation operators for the oscillators, and the $g_k$ describe the interaction strength between the two-level system and the environment modes. This system-environment Hamiltonian is widely used to study, for instance, spontaneous emission \cite{Scullybook}. Note that this Hamiltonian is the same as 
	$$ H = \frac{\varepsilon}{2}\sigma_z + \sum_k \omega_k b_k^\dagger b_k +  \sigma_x \sum_k (g_k^* b_k + g_k b_k^\dagger),$$
	except that the non-rotating wave approximation terms have been dropped. To calculate the filter function, we identify $F_1 = \sigma^+$, $B_1 = \sum_k g_k^* b_k$, $F_2 = \sigma^-$ and $B_2 = \sum_k g_k b_k^\dagger$. Let us first calculate the environment correlation functions. We find that 
	$$ C_{12}(t') = \sum_{k,k'} g_{k'} g_k^* e^{-i\omega_k t'}\langle b_k b_{k'}^\dagger \rangle_B. $$
	At zero temperature, all the oscillators are in their ground state, and we get $C_{12}(t') = \sum_k |g_k|^2 e^{-i\omega_k t'}$, thus allowing us to identify $f_{12}(\omega_k,t') = e^{-i\omega_k t'}$. On the other hand, it is straightforward to show that $C_{11}(t') = C_{22}(t') = 0$, and $C_{21}(t') = 0$ at zero temperature as well. We can thus write 
	\begin{align*}
			Q(\omega,\tau) = &\frac{2}{\tau} \,\re \biggl[ \int_0^\tau dt \int_0^{t} dt' \biggl( e^{-i\omega t'} \notag \\
			&\times \tr \bigl\lbrace P_{\psi^\perp} \widetilde{F}_2(t - t')\rho_S(0) \widetilde{F}_1(t) \bigr\rbrace \biggr) \biggr].
			\end{align*}	
	The state initially prepared is the excited state which we denote by $\ket{\uparrow}$. It follows that $P_{\psi^\perp} = \ket{\downarrow}\bra{\downarrow}$, where $\sigma_z\ket{\downarrow} = -\ket{\downarrow}$. Using $\widetilde{F}_1(t) = \sigma^+ e^{i\varepsilon t}$ and $\widetilde{F}_2(t - t') = \sigma^- e^{-i\varepsilon (t - t')}$, we find that 
			\begin{align*}
			Q(\omega,\tau) = &\frac{2}{\tau} \int_0^\tau dt \int_0^{t} dt' \cos[(\varepsilon - \omega_k)t'].
			\end{align*}	
			Performing the integrals, we end up with 
			$$ Q(\omega,\tau) = \tau \:\text{sinc}^2 \left[ \frac{(\varepsilon - \omega)\tau}{2} \right], $$
			which is the usual filter function \cite{KurizkiNature2000,KoshinoPhysRep2005}. Thus, our formalism reproduces the well-known sinc-squared function for the case where we study the Zeno to anti-Zeno transition in the context of population decay. In particular, if $\tau$ is small, then we obtain $\Gamma(\tau) \approx \tau \int_0^\infty d\omega \, J(\omega)$, thus putting us in the Zeno regime. However, for larger $\tau$, the overlap between $J(\omega)$ and $Q(\omega,\tau)$ can increase, leading to the anti-Zeno effect. 
			
			\subsection{The pure dephasing model}
			
			Let us now consider the system environment model specified by the Hamiltonian
			$$ H = \frac{\varepsilon}{2}\sigma_z + \sum_k \omega_k b_k^\dagger b_k + \sigma_z  \sum_k (g_k^* b_k + g_k b_k^\dagger). $$
			A two-level system with level spacing $\varepsilon$ is interacting with an environment that is again modeled as a collection of harmonic oscillators. However, now there is no population decay. Instead, the system undergoes dephasing only, which is the reason why this model is known as the pure dephasing model \cite{ChaudhryPRA2014zeno}. Once again, we start by calculating the environment correlation function. It is straightforward to show that 
			$$ C(t') = \sum_k |g_k|^2 \left[\cos (\omega_k t') \coth\left( \frac{\beta \omega_k}{2}\right)  - i \sin(\omega_k t')\right], $$
			where we have assumed $\rho_B = e^{-\beta H_B}/Z_B$, that is, the environment is in the standard thermal equilibrium state. Thus, we can read off that $f(\omega_k,t') = \cos (\omega_k t') \coth\left( \frac{\beta \omega_k}{2}\right)  - i \sin(\omega_k t')$. We consider the system state that we start off with to be $\ket{\uparrow_x} = \frac{1}{\sqrt{2}}(\ket{\uparrow} + \ket{\downarrow})$, where, as before, $\sigma_z\ket{\uparrow} = \ket{\uparrow}$ and $\sigma_z\ket{\downarrow} = -\ket{\downarrow}$. We repeatedly measure to check if the state is still $\ket{\uparrow_x}$ or not with time interval $\tau$. Now,
			$$ \tr\lbrace P_{\psi^\perp} \sigma_z \ket{\uparrow_x}\bra{\uparrow_x}\sigma_z \rbrace = 1. $$
			Therefore, 
			$$ Q(\omega_k,\tau) = \frac{2}{\tau}\int_0^\tau dt \int_0^t dt' \cos(\omega_k t') \coth \left( \frac{\beta \omega_k}{2} \right). $$
			Performing the integrals and simplifying, we obtain
			\begin{equation*}
			\Gamma(\tau) = \sum_k |g_k|^2 \frac{2}{\tau} \coth \left( \frac{\beta \omega_k}{2} \right) \left[ \frac{1 - \cos(\omega_k \tau)}{\omega_k^2}\right]. 
			\end{equation*}
			On the other hand, the pure dephasing model can be solved exactly \cite{ChaudhryPRA2014zeno}. Using the exact solution, it can be shown that 
			$$ \Gamma(\tau) = -\frac{1}{\tau} \ln \left[ 1 - \frac{1}{2}(1 - e^{-\gamma(\tau)})\right], $$
			where 
			$$ \gamma(\tau) = \sum_k |g_k|^2 \frac{4}{\omega_k^2} [1 - \cos(\omega_k \tau)] \coth \left( \frac{\beta \omega_k}{2} \right). $$
			For weak coupling, however, $1 - e^{-\gamma(\tau)} \approx \gamma(\tau)$, leading to 
			\begin{align*}			
			\Gamma(\tau) &\approx -\frac{1}{\tau} \ln\left[ 1 - \frac{1}{2} \gamma(\tau) \right] \\
			&\approx \frac{1}{2\tau}\gamma(\tau) \\
			&= \sum_k |g_k|^2 \frac{2}{\tau} \coth \left( \frac{\beta \omega_k}{2} \right) \left[ \frac{1 - \cos(\omega_k \tau)}{\omega_k^2}\right],
			\end{align*}
			which agrees with the result obtained using the formalism that we have developed. 
			
			\subsection{The general spin-boson model}
			
			We now consider the more general system-environment model given by the Hamiltonian 
			$$ H = \frac{\varepsilon}{2}\sigma_z + \frac{\Delta}{2}\sigma_x + \sum_k \omega_k b_k^\dagger b_k + \sigma_z \sum_k (g_k^* b_k + g_k b_k^\dagger),$$
			where $\Delta$ can be understood as the tunneling amplitude for the system, and the rest of the parameters are defined as before. This is the well-known spin-boson model \cite{LeggettRMP1987,Weissbook,BPbook}, which can be considered as an extension of the previous two cases in that we can now generally have both population decay and dephasing taking place. We revert to the usual dephasing model by setting $\Delta = 0$, while setting $\varepsilon = 0$ leads to the population decay Hamiltonian (with the non-rotating wave approximation terms now included) after rotation about the $y$-axis. Experimentally, such a model can be realized, for instance, using superconducting qubits \cite{ClarkeNature2008, YouNature2011,Slichterarxiv2015} and the properties of the environment can be appropriately tuned as well \cite{HurPRB2012}. We start from the general system initial state 
			$$ \ket{\psi} = \cos\left(\frac{\theta}{2}\right)\ket{\uparrow} + e^{i\phi}\sin\left(\frac{\theta}{2}\right)\ket{\downarrow}, $$
			where the states $\ket{\uparrow}$ and $\ket{\downarrow}$ are defined as before, and $\theta$ and $\phi$ are parameters that characterize the state preparation. Measurements, with time interval $\tau$, are now carried out to check if the system is still in this state or not.
			To evaluate $Q(\omega,\tau)$, we first need to find the environment correlation function. We again consider the environment to be in a thermal equilibrium state. Consequently, as in the pure dephasing model, $f(\omega,t') = \cos(\omega t') \coth(\beta \omega/2) - i \sin(\omega t')$. Next, we find $\widetilde{F}(t)$. This is done using the standard commutation relation $[\sigma_k, \sigma_l] = 2i\varepsilon_{klm}\sigma_m$.  A straightforward application of the Baker-Hausdorff lemma \cite{Puribook} shows that 
			\begin{equation*}
			\widetilde{F}(t) = a_x(t) \sigma_x + a_y(t) \sigma_y + a_z(t)\sigma_z, 
			\end{equation*}
			with 
			\begin{align*}
			a_x(t) &= \frac{2\varepsilon \Delta}{\Omega^2} \sin^2 \left( \frac{\Omega t}{2} \right), \\
			a_y(t) &= \frac{\Delta}{\Omega} \sin(\Omega t), \\
			a_z(t) &= 1 - \frac{2\Delta^2}{\Omega^2} \sin^2 \left( \frac{\Omega t}{2} \right), 
			\end{align*}
			with $\Omega^2 = \varepsilon^2 + \Delta^2$. Now, given $\ket{\psi}$, we can deduce that $P_{\psi^\perp} = \ket{\psi^\perp}\bra{\psi^\perp}$ 		with 
			$$ \ket{\psi^\perp} = \sin\left(\frac{\theta}{2}\right)\ket{\uparrow} - e^{i\phi}\cos\left(\frac{\theta}{2}\right)\ket{\downarrow}. $$
			Using this, we then find
			$$ \opav{\psi}{\widetilde{F}(t)}{\psi^\perp} = r_1(t) + ir_2(t), $$
			where 
			\begin{align*}
			r_1(t) &= -a_x(t) \cos \phi \cos \theta - a_y(t) \sin \phi \cos \theta + a_z(t) \sin \theta, \\
			r_2(t) &= a_y(t) \cos \phi - a_x(t) \sin \phi.
			\end{align*}
			Similarly,
			$$ \opav{\psi^\perp}{\widetilde{F}(t - t')}{\psi} = r_1(t - t') - ir_2(t - t'). $$
			Thus, the generalized filter function becomes 
			\begin{equation}
			Q(\omega,\tau) = \frac{2}{\tau}\left\lbrace \coth \left(\frac{\beta \omega}{2}\right) D_1(\omega,\tau)  + D_2(\omega,\tau)\right\rbrace,
			\end{equation}
			with 
			\begin{align*}
			&D_1(\omega,\tau) = \notag \\
			&\int_0^\tau dt \int_0^t dt' \cos (\omega t') [r_1(t - t') r_1(t) + r_2(t - t')r_2(t)], \\
			&D_2(\omega,\tau) = \notag \\
			&\int_0^\tau dt \int_0^t dt' \sin(\omega t') [r_1(t - t')r_2(t) - r_1(t) r_2(t - t')].
			\end{align*}
			
			We next evaluate $D_1(\omega,\tau)$ and $D_2(\omega,\tau)$. Although the integrals can be done in a straightforward manner, the final analytical results for arbitrary $\theta$ and $\phi$ are, unfortunately, very long and not very illuminating. Instead, let us choose $\theta = \pi/2$ and $\phi = 0$. This choice of state has the advantage that we can then compare our results with the already well-known cases of the population decay model and the pure dephasing model. In particular, the choice $\Delta = 0$ then corresponds to the pure dephasing model (in the weak coupling limit). On the other hand, $\varepsilon = 0$ corresponds to (almost) the population decay model since we can rotate both the system-environment Hamiltonian and the state that we are measuring about the $y$-axis. The only difference is that we now have additional non-rotating wave approximation terms, but we expect these additional terms to not play a role in the weak system-environment coupling regime that we are in. It must be emphasized that we are now no longer restricted to these two models only. Rather, by varying the values of $\theta$, $\phi$, $\varepsilon$, and $\Delta$, we can explore regimes where both dephasing and relaxation play a role in the quantum Zeno and anti-Zeno effects. 
			
\begin{figure}
\centering
{\includegraphics[scale = 0.5]{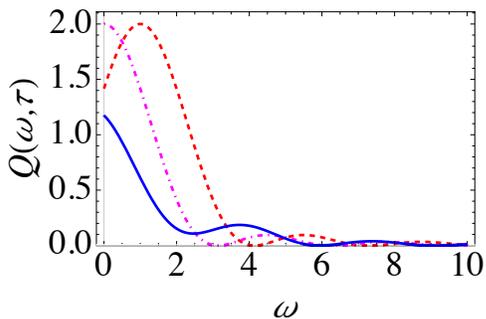}}
\caption{(Color online) Behavior of $Q(\omega,\tau)$ (at zero temperature) as a function of $\omega$ for $\tau = 2$ with $\Delta = 1$ and $\varepsilon = 0$ (dashed, red curve), $\Delta = 0$ and $\varepsilon = 1$ (dot-dashed, magenta curve), and $\Delta = 1$ and $\varepsilon = 2$ (solid, blue curve). Here, we have set $\theta = \pi/2$ and $\phi = 0$. Throughout, we use dimensionless units with $\hbar = 1$.}
\label{filterfunctiongraphtau2}
\end{figure}

\begin{figure}
\centering
{\includegraphics[scale = 0.5]{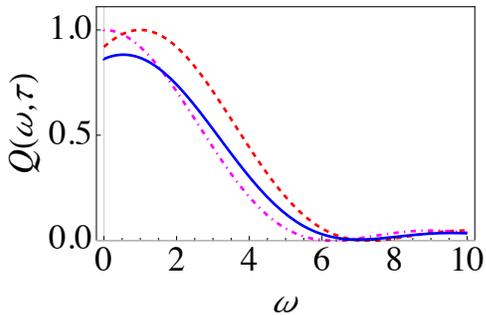}}
\caption{(Color online) Same as Fig.~\ref{filterfunctiongraphtau2}, except that now $\tau = 1$.}
\label{filterfunctiongraphtau1}
\end{figure}

			With $\theta = \pi/2$ and $\phi = 0$, the calculation for $D_1(\omega,\tau)$ and $D_2(\omega,\tau)$ becomes less laborious since $r_1(t)$ and $r_2(t)$ simplify greatly. Analytical expressions for $D_1(\omega,\tau)$ and $D_2(\omega,\tau)$ are given in the Appendix. A few points are in order. First, if $\Delta = 0$, then $D_2(\omega,\tau) = 0$, while $D_1(\omega,\tau) = [1- \cos(\omega \tau)]/\omega^2$, which, as expected, leads back to the filter function for the pure dephasing model. Similarly, for zero temperature with $\varepsilon = 0$, $D_1(\omega,\tau)$ and $D_2(\omega,\tau)$ simplify such that we get back the filter function for the population decay model. For intermediate values of $\varepsilon$ and $\Delta$, however, the filter function is very different. This is illustrated in Figs.~\ref{filterfunctiongraphtau2} and \ref{filterfunctiongraphtau1}, where $Q(\omega,\tau)$ has been plotted as a function of $\omega$ for two different values of $\tau$. Note that throughout this work, we are using dimensionless units with $\hbar = 1$. The dashed, red curve is the filter function for the population decay model with $\varepsilon = 0$ and $\Delta = 1$, and is thus consequently peaked at $\omega = 1$ in both of the figures. The dot-dashed, magenta curve is the filter function for the pure dephasing model (with $\varepsilon = 1$ and $\Delta = 0$), while the solid, blue curve shows the intermediate case with $\varepsilon = 2$ and $\Delta = 1$. Clearly, if the value of $\tau$ is not too small, the filter function for the intermediate case resembles neither the population decay model nor the pure dephasing model [see Fig.~\ref{filterfunctiongraphtau2}]. This means that for an arbitrary spectral density of the environment, the value of $\Gamma(\tau)$ is expected to be quite different with both $\Delta$ and $\varepsilon$ non-zero as compared to the population decay and pure dephasing cases. On the other hand, as $\tau$ becomes smaller, we must have that $\Gamma(\tau) \rightarrow 0$, and thus the filter functions must start to resemble each other more for smaller $\tau$. This is entirely consistent with what we see in Fig.~\ref{filterfunctiongraphtau1}.

\begin{figure}
\centering
{\includegraphics[scale = 0.75]{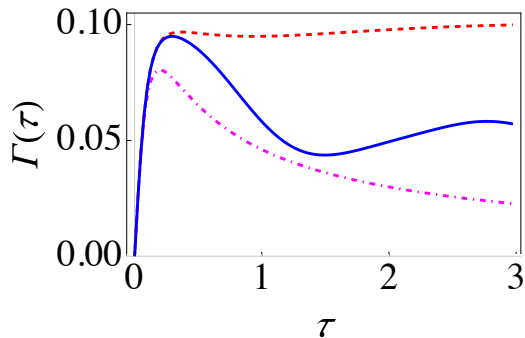}}
\caption{(Color online) Graph of $\Gamma(\tau)$ (at zero temperature) as a function of $\tau$ for $\theta = \pi/2$ and $\phi = 0$ with $\Delta = 2$ and $\varepsilon = 0$ (dashed, red curve), $\Delta = 0$ and $\varepsilon = 2$ (dot-dashed, magenta curve), and $\Delta = 2$ and $\varepsilon = 2$ (solid, blue curve) for an Ohmic environment $(s = 1)$. We have set $G = 0.01$ and $\omega_c = 10$.}
\label{gammathetapiby2ohmic}
\end{figure}

With the different behavior of the filter function, we expect that when there is both dephasing and decay, the effective decay rate will be quite different compared to either of the aforementioned cases, and thus the Zeno and anti-Zeno behavior is expected to be considerably modified. Let us now explicitly examine this claim for the case of sub-Ohmic, Ohmic and super-Ohmic environments. To this end, we introduce the spectral density as 
\begin{equation*}
J(\omega) = G \omega^s \omega_c^{1 - s} e^{-\omega/\omega_c}. 
\end{equation*}
Here the parameter $s$ characterizes the Ohmicity of the environment. Namely, $s = 1$ corresponds to an Ohmic environment, $s > 1$ gives a super-Ohmic environment, while $s < 1$ corresponds to a sub-Ohmic environment. $G$ is a dimensionless parameter characterizing the system-environment coupling strength, and we have introduced an exponential cutoff function with a cutoff frequency $\omega_c$. Let us first investigate an Ohmic environment. In Fig.~\ref{gammathetapiby2ohmic}, we illustrate the t  of the effective decay rate $\Gamma(\tau)$ as a function of the measurement interval $\tau$, clearly showing the very different behavior of $\Gamma(\tau)$ when both population decay and dephasing are present. For the population decay case (the dashed, red curve), with the chosen values of the system-environment parameters, $\Gamma(\tau)$ by and large decreases as $\tau$ decreases. Thus, we are in the Zeno regime. On the other hand, the pure dephasing case (dot-dashed, magenta curve) displays a distinct Zeno regime and an anti-Zeno regime. For small values of $\tau$, $\Gamma(\tau)$ decreases as $\tau$ is decreased, meaning that a shorter measurement interval helps to protect the quantum state, thus putting us in the Zeno regime. For larger values of $\tau$, however, as the $\tau$ is decreased the opposite happens, namely, $\Gamma(\tau)$ increases as $\tau$ decreases, thus indicating the anti-Zeno regime. Now, when both dephasing and population decay take place (solid, blue curve), the behavior of $\Gamma(\tau)$ is considerably different. Besides the quantitative differences in the values of $\Gamma(\tau)$, the effective decay rate displays qualitatively different behavior in the sense that we now have clearly distinct multiple Zeno and anti-Zeno regions. This is evident from the fact that $\Gamma(\tau)$ displays multiple extrema, meaning that sometimes decreasing the measurement interval reduces the decay rate, while sometimes the opposite happens. 

\begin{figure}
\centering
{\includegraphics[scale = 0.75]{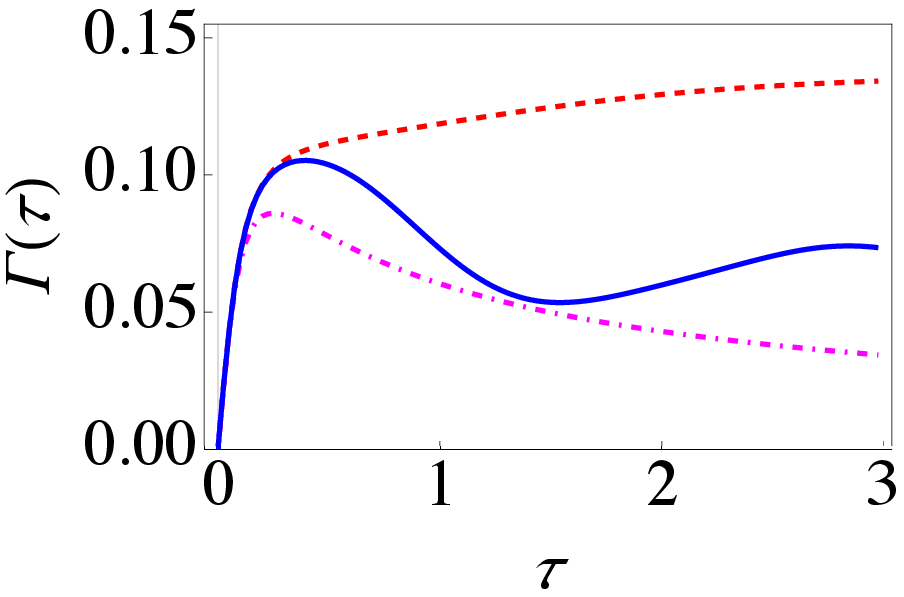}}
\caption{(Color online) Same as Fig.~\ref{gammathetapiby2ohmic}, except that now we have a sub-Ohmic environment with $s = 0.8$.}
\label{gammathetapiby2subohmic}
\end{figure}

\begin{figure}
\centering
{\includegraphics[scale = 0.75]{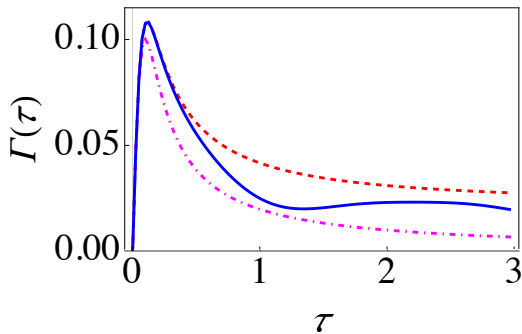}}
\caption{(Color online) Same as Fig.~\ref{gammathetapiby2ohmic}, except that now we have a super-Ohmic environment with $s = 2$.}
\label{gammathetapiby2superohmic}
\end{figure}

The effective decay rate, as we have emphasized, depends on the overlap of the spectral density and the generalized filter function. Thus, we expect that changing the environment, and in particular the Ohmicity parameter, should alter the decay rate, at least quantitatively. In Fig.~\ref{gammathetapiby2subohmic}, we have calculated the decay rate with a similar set of parameters as was done with the Ohmic environment [see Fig.~\ref{gammathetapiby2ohmic}]. The only difference is that we are now considering a sub-Ohmic environment with $s = 0.8$. Once again, while the population decay case exhibits the QZE, and the pure dephasing case exhibits both the QZE and the QAZE, the more general case displays multiple transitions between the QZE and the QAZE. We have also examined a super-Ohmic environment [see Fig.~\ref{gammathetapiby2superohmic}] with $s = 2$. Now the population decay case also exhibits both the QZE and the QAZE. With both dephasing and population decay present, we again have multiple Zeno to anti-Zeno transitions, but these transitions are less clear cut as compared to what we observed in the previous cases. These results illustrate the importance of the type of environment in determing the QZE-QAZE crossover behavior. 

\begin{figure}
\centering
{\includegraphics[scale = 0.75]{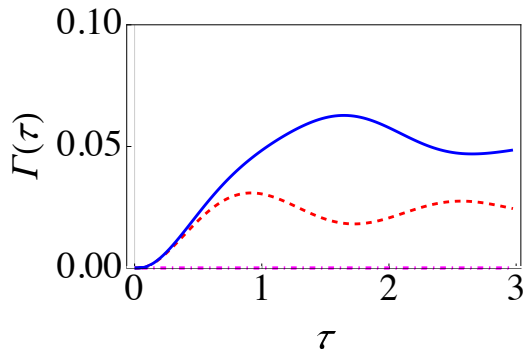}}
\caption{(Color online) Graph of $\Gamma(\tau)$ (at zero temperature) as a function of $\tau$ for $\theta = 0$ and $\phi = 0$ with $\Delta = 2$ and $\varepsilon = 0$ (dashed, red curve), $\Delta = 0$ and $\varepsilon = 2$ (dot-dashed, magenta curve), and $\Delta = 2$ and $\varepsilon = 2$ (solid, blue curve) for an Ohmic environment $(s = 1)$. We have set $G = 0.01$ and $\omega_c = 10$.}
\label{gammathetazeroohmic}
\end{figure}

Before moving on, let us also investigate the Zeno and anti-Zeno effects for a different state preparation. Namely, we choose $\theta = 0$ and $\phi = 0$. Once again we find $D_1(\omega,\tau)$ and $D_2(\omega,\tau)$ and use these to find the effective decay rate. Exact expressions for $D_1(\omega,\tau)$ and $D_2(\omega,\tau)$ can be found in a similar way as before (refer to the Appendix). With these expressions in hand, we can investigate the behavior of $\Gamma(\tau)$. In Fig.~\ref{gammathetazeroohmic}, we have shown the behavior of $\Gamma(\tau)$ for the population decay Hamiltonian (dashed, red curve), the pure dephasing case (dot-dashed, magenta curve), and the intermediate case (solid, blue curve) with an Ohmic environment. For pure dephasing,  $\Gamma(\tau)$ remains zero. This makes sense since the state that we are repeatedly preparing, namely $\ket{\uparrow}$, does not evolve under the action of the pure dephasing Hamiltonian. On the other hand, the population decay Hamiltonian leads to both decay and dephasing, since, after rotation about the $y$-axis, the state that is repeatedly prepared is a superposition of the $\ket{\uparrow}$ and $\ket{\downarrow}$ states. Although both the population decay Hamiltonian and the intermediate case display multiple Zeno to anti-Zeno transitions, the actual value of the decay rate, in general, is considerably different for the two cases. These two cases also differ in the values of $\tau$ for which the transitions take place. Once again, it is clear that using the usual sinc-squared filter function to analyze a system undergoing both dephasing and decay would be incorrect.

\subsection{Large spin}

\begin{figure}
\centering
{\includegraphics[scale = 0.75]{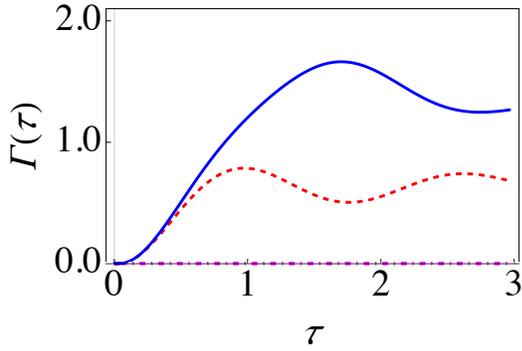}}
\caption{(Color online) Behavior of $\Gamma(\tau)$ (at zero temperature) as a function of $\tau$ for the initial state $\ket{j}$ with $\Delta = 2$ and $\varepsilon = 0$ (dashed, red curve), $\Delta = 0$ and $\varepsilon = 2$ (dot-dashed, magenta curve), and $\Delta = 2$ and $\varepsilon = 2$ (solid, blue curve) for a sub-Ohmic environment $(s = 0.8)$. We have set $G = 0.01$ and $\omega_c = 10$, and the number of particles is taken to be $20$.}
\label{gammathetazerosubohmiclarge}
\end{figure}

Let us now consider $N_s$ two-level systems interacting collectively with a common environment. This situation can be modeled via the system-environment Hamiltonian
$$ H = \varepsilon J_z + \Delta J_x + \sum_k \omega_k b_k^\dagger b_k + 2J_z \sum_k (g_k^* b_k + g_k^\dagger b_k^\dagger),$$
where $\varepsilon$ and $\Delta$ are the level spacing and tunneling amplitude respectively for each two-level system, and $J_x$ and $J_z$ are the standard angular momentum operators. This Hamiltonian can be considered to be a generalization of the usual spin-boson model to a large spin $j = N_s/2$ \cite{ChaudhryPRA2014zeno,VorrathPRL2005,KurizkiPRL2011}. Physical realizations include a two-component Bose-Einstein condensate \cite{GrossNature2010,RiedelNature2010} that interacts with a thermal reservoir via collisions \cite{KurizkiPRL2011}. Once again, we assume that the system-environment coupling is weak, so that our formalism applies. The initial state is chosen to be $\ket{j}$ such that $J_z\ket{j} = j\ket{j}$, and we repeatedly check, with time interval $\tau$, whether the system state is still $\ket{j}$ or not. We note that we can easily deal with some other choice of initial state as well. As before, our task is to calculate $Q(\omega,\tau)$. We begin by calculating $\widetilde{F}(t) = e^{iH_S t} (2J_z) e^{-iH_S t}$. Using the standard commutation relations, $[J_k,J_l] = i\varepsilon_{klm} J_m$, and the Baker-Hausdorff lemma, we find that
$$ \widetilde{F}(t) = 2[a_x(t) J_x + a_y(t) J_y + a_z(t) J_z], $$
with $a_x(t)$, $a_y(t)$ and $a_z(t)$ as defined before. Now, $P_{\psi^\perp} = \sum_{m = -j}^{j - 1} \ket{m}\bra{m}$. We then find that 
$$ \tr \lbrace P_{\psi^\perp} \widetilde{F}(t - t')\rho_S(0)\widetilde{F}(t) \rbrace =  \sum_{m = -j}^{j - 1} \opav{m}{\tilde{F}(t - t')}{j} \opav{j}{\tilde{F}}{m}.$$ 
By observing that $\opav{j}{\tilde{F}(t)}{m} = \sqrt{2j}\delta_{m,j-1} [a_x(t) - ia_y(t)]$, we find that now
\begin{equation}
Q(\omega,\tau) = (2j)\frac{2}{\tau} \left\lbrace \coth\left(\frac{\beta \omega}{2} \right) D_1(\omega,\tau) + D_2(\omega,\tau)\right\rbrace,
\end{equation}
with 
\begin{align*}
D_1(\omega,\tau) = \int_0^\tau dt \int_0^t dt' \cos (\omega t') &\lbrace a_x(t) a_x(t - t') + \notag \\
			&a_y(t) a_y(t - t')\rbrace,
\end{align*}
\begin{align*}
D_2(\omega,\tau) = \int_0^\tau dt \int_0^t dt' \sin (\omega t') &\lbrace a_x(t) a_y(t - t') - \notag \\
			&a_y(t) a_x(t - t')\rbrace.
\end{align*}
This agrees with what we found before for $j = 1/2$ with $\theta = 0$ and $\phi = 0$, meaning that for the large spin-boson model, $\Gamma(\tau)$ is amplified depending on the number of two-level systems collectively coupled to the environment. In other words, $\Gamma(\tau)$ for $N_s$ particles is simply $N_s$ times the decay rate for a single particle. For illustration purposes, we have plotted $\Gamma(\tau)$ as a function of the measurement interval $\tau$ in Fig.~\ref{gammathetazerosubohmiclarge} for $N_s = 20$ and a sub-Ohmic environment. The important point to note here is that although the qualitative behavior of $\Gamma(\tau)$ is similar to that for the single two-level system case, the amplification of $\Gamma(\tau)$ makes the use of the correct filter function even more critical. For instance, for the population decay Hamiltonian in Fig.~\ref{gammathetazerosubohmiclarge} with $\Delta = 2$ and $\varepsilon = 0$, the survival probability after five measurements with time interval $\tau = 1$ is approximately 0.02, while the survival probability for the more general Hamiltonian with $\Delta = 2$ and $\varepsilon = 2$ after five measurements with the same time interval $\tau = 1$ is ten times smaller.

\begin{figure}
\centering
{\includegraphics[scale = 0.75]{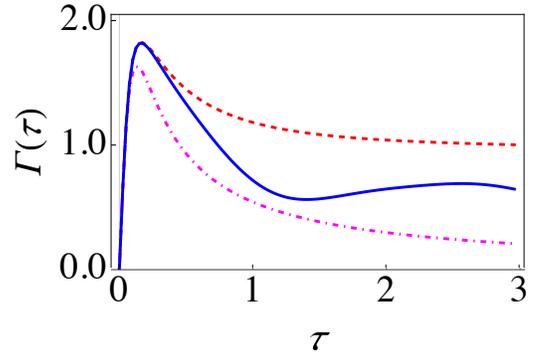}}
\caption{(Color online) Same as Fig.~\ref{gammathetazerosubohmiclarge}, except that now the state that we are repeatedly preparing is $\ket{\psi_c}$ and the environment is super-Ohmic with $s = 1.5$.}
\label{gammathetapiby2superohmiclarge}
\end{figure}

Suppose now that we prepare the state of each two-level system in a coherent superposition such that the total state is $\ket{\psi_c}$ where $J_x\ket{\psi_c} = j\ket{\psi_c}$. To investigate the Zeno and anti-Zeno effects now, it is best to transform to a rotated frame, that is, we rotate both the system-environment Hamiltonian and the state that we are repeatedly preparing. In other words, we use the fact that 
\begin{align*}
s(\tau) &= \tr_{S,B}[(\ket{\psi_c}\bra{\psi_c}) U(\tau) (\ket{\psi_c}\bra{\psi_c}) U^\dagger (\tau) ] = \notag \\
&= \tr_{S,B} [(\ket{j}\bra{j}) e^{-iH_R \tau} (\ket{j}\bra{j}) e^{iH_R \tau}],
\end{align*}
where the transformed Hamiltonian is 
\begin{align*}
H_R = \varepsilon_r J_z + \Delta_r J_x + \sum_k \omega_k b_k^\dagger b_k - 2J_x \sum_k (g_k^* b_k + g_k b_k^\dagger),
\end{align*}
with $\varepsilon_r = \Delta$ and $\Delta_r = -\varepsilon$. We then find that 
$$\tilde{F}(t) = -2[b_x(t) J_x + b_y(t) J_y + b_z(t) J_z], $$
with 
\begin{align*}
b_x(t) = 1 - \frac{2\varepsilon_r^2}{\Omega_r^2} \sin^2 \left(\frac{\Omega_r t}{2}\right), \\
b_y(t) = -\frac{\varepsilon_r}{\Omega_r} \sin(\Omega_r t), \\
b_z(t) = \frac{2\varepsilon_r \Delta_r}{\Omega_r^2} \sin^2 \left(\frac{\Omega_r t}{2}\right),
\end{align*}
where $\Omega_r = \sqrt{\varepsilon_r^2 + \Delta_r^2}$. The rest of the calculation proceeds in a very similar way to what we did with the state $\ket{j}$ and leads to 
\begin{equation*}
Q(\omega,\tau) = (2j)\frac{2}{\tau} \left\lbrace \coth\left(\frac{\beta \omega}{2} \right) D_1^R(\omega,\tau) + D_2^R(\omega,\tau)\right\rbrace,
\end{equation*}
with 
\begin{align*}
D_1^R(\omega,\tau) = \int_0^\tau dt \int_0^t dt' \cos (\omega t') &\lbrace b_x(t) b_x(t - t') + \notag \\
			&b_y(t) b_y(t - t')\rbrace,
\end{align*}
\begin{align*}
D_2^R(\omega,\tau) = \int_0^\tau dt \int_0^t dt' \sin (\omega t') &\lbrace b_x(t) b_y(t - t') - \notag \\
			&b_y(t) b_x(t - t')\rbrace.
\end{align*}
Now, it is easy to show that $b_x(t) = a_z(t)$ and $b_y(t) = -a_y(t)$. Thus, we reduce to the results obtain for a single two-level system when we repeatedly prepare the state $\ket{\uparrow_x}$. However, now we can consider an arbitrary value of $N_s$. We again find that $\Gamma(\tau)$ is amplified depending on the number of two-level systems coupled to the environment. Quantitative results are presented in Fig.~\ref{gammathetapiby2superohmiclarge} for $N_s = 20$ with a super-Ohmic environment $(s = 1.5)$. Again, due to the amplification of the effective decay rate, it becomes very important to use the correct filter function. 

\section{Conclusion}

In conclusion, we have derived an expression for the effective decay rate of a quantum state in the presence of repeated measurements which is valid when the system-environment coupling is weak. This expression implies that the effective decay rate of the quantum state depends on the overlap of the spectral density of the environment and a generalized filter function that itself depends on the system state that is repeatedly being prepared, the measurement interval, the system parameters, the system-environment coupling, and the environment correlation function. We have shown that our formalism for calculating the effective decay rate reproduces the well-known results for the population decay model and the pure dephasing model. Thereafter, we demonstrated that our formalism allows us to study the Zeno and anti-Zeno effects for the spin-boson model in a rigorous fashion under the assumption that the system-environment coupling strength is weak. We have found qualitative and quantitative differences in the behavior of the decay rate as a function of the measurement interval when both decay and dephasing are present as compared to the relatively simpler population decay and pure dephasing models. Finally, by considering many two-level systems coupled collectively to a common environment, we have observed that the decay rate is amplified depending on the number of two-level systems. Consequently, it is even more crucial to use the correct filter function to evaluate the effective decay rate. Experimental implementations of the ideas presented in this paper are expected to be important for measurement-based quantum control.

\begin{acknowledgements}

Support from the LUMS startup grant is gratefully acknowledged. 

\end{acknowledgements}

\newpage

\onecolumngrid

\appendix

\section{Calculating the integrals}

In this appendix, we present the analytical expressions for $D_1(\omega,\tau)$ and $D_2(\omega,\tau)$ for a single two-level system. For $\theta = \pi/2$ and $\phi = 0$, we find that $D_1(\omega,\tau) = X_1/X_2$, and $D_2(\omega,\tau) = X_3/X_4$, where 
\begin{align*}
X_1 = &3 \Delta ^4 \omega ^4 + 4 \Omega ^2 \cos (\omega \tau) \left[\varepsilon^2 (\varepsilon^2 - \omega^2)(\omega^2 - \Omega^2)-\Delta ^2 \omega ^2 \left(\Delta ^2+\omega ^2\right) \cos (\Omega \tau)\right] + \\
			&\Delta ^2 \omega  \left[4 \Omega ^3 \left(\varepsilon^2 - 2 \omega ^2\right) \sin (\omega \tau) \sin (\Omega \tau)-\omega  \varepsilon^2 (\omega^2 -\Omega^2 )(\cos (2\Omega \tau)-4 \cos (\Omega \tau))\right]- \\
			&8 \Omega ^6 \left(\Delta ^2+\omega ^2\right)-3 \Delta ^2 \omega ^2 \Omega ^2 \left(\Delta ^2+\omega ^2\right)+\Omega ^4 \left(4 \Delta ^4+15 \Delta ^2 \omega ^2+4 \omega ^4\right)+4 \Omega ^8, \\
X_2 &= 4 \Omega ^4 \omega^2 \left(\omega ^2-  \Omega ^2\right)^2, \\ 
X_3 = &\Delta  \biggl\lbrace \omega  \Omega (\varepsilon^2 - \Delta ^2-\omega ^2) \sin (\omega \tau) \sin (\Omega \tau  )+ \cos (\omega \tau  ) [\varepsilon^2 (\Omega^2 - \omega^2)- (\Omega ^2 \left(\Delta ^2+\omega ^2)+\Delta ^2 \omega ^2-\Omega ^4\right) \cos (\Omega \tau  )] + \\
&\Omega ^2 (\Delta ^2+\omega ^2)+\Delta ^2 \omega ^2 + \varepsilon^2 (\omega^2 - \Omega^2) \cos (\Omega \tau)-\Omega ^4 \biggr\rbrace, \\
X_4 &= \omega  \Omega ^2 \left(\omega ^2-\Omega ^2\right)^2.
\end{align*}
On the other hand, for $\theta = 0$ and $\phi = 0$, we find that 
\begin{align*}
X_1 = &\Delta ^2 \biggl\lbrace \omega ^2 \Omega ^2 \left(\omega ^2+3 \Omega ^2\right)-4 \omega  \Omega ^3 \left(\omega ^2+\varepsilon ^2\right) \sin (\omega \tau) \sin (\Omega \tau)+4 \Omega ^2 \cos (\omega \tau) \left[\varepsilon ^2 (\omega^2 -\Omega^2 )-\omega ^2 \left(\Omega ^2+\varepsilon ^2\right) \cos (\Omega \tau)\right] + \notag \\
&\omega ^2 (\Omega^2 -\omega^2 )\left[\Delta^2 \cos (2 \Omega \tau)+4 \varepsilon ^2 \cos (\Omega \tau)\right]+\varepsilon ^2 \left(3 \omega ^4-3 \omega ^2 \Omega ^2+4 \Omega ^4\right)\biggr\rbrace, \\
X_2 &= 4 \Omega ^4 \omega^2 \left(\omega ^2-\Omega ^2\right)^2, \\
X_3 &= 4 \Delta ^2 \varepsilon  \left\lbrace\omega  \cos \left(\frac{\omega \tau}{2}\right) \sin \left(\frac{\Omega \tau}{2}\right)-\Omega  \sin \left(\frac{\omega \tau}{2}\right) \cos \left(\frac{\Omega \tau}{2}\right)\right\rbrace^2, \\
X_4 &= \omega  \Omega ^2 \left(\omega ^2-\Omega ^2\right)^2.
\end{align*}

\twocolumngrid


\begin{thebibliography}{99}
\bibitem{noteoncorrelations} We are ignoring the build-up of the system-environment correlations as the system and environment evolve together. That is, we are neglecting any disturbance to the environment induced by measuring the system. This is valid in the weak system-environment coupling regime that we will be dealing with \cite{ChaudhryPRA2014zeno}.
\end{thebibliography}
\end{document}